\documentclass[preprint,journal]{vgtc}       % preprint (journal style)

\ifpdf%                                % if we use pdflatex
  \pdfoutput=1\relax                   % create PDFs from pdfLaTeX
  \usepackage{graphicx}                % allow us to embed graphics files
  \DeclareGraphicsExtensions{.pdf,.png,.jpg,.jpeg} % for pdflatex we expect .pdf, .png, or .jpg files
\else%                                 % else we use pure latex
  \usepackage{graphicx}                % allow us to embed graphics files
  \DeclareGraphicsExtensions{.eps}     % for pure latex we expect eps files
\fi%

\graphicspath{{figures/}{pictures/}{images/}{./}} % where to search for the images

\usepackage{microtype}                 % use micro-typography (slightly more compact, better to read)
\PassOptionsToPackage{warn}{textcomp}  % to address font issues with \textrightarrow
\usepackage{textcomp}                  % use better special symbols
\usepackage{mathptmx}                  % use matching math font
\usepackage{times}                     % we use Times as the main font
\usepackage{fontawesome}
 
\usepackage{balance}  % a nicer typewriter font
\usepackage{cite}                      % needed to automatically sort the references
\usepackage{tabu}                      % only used for the table example
\usepackage{booktabs}
\usepackage{color,soul}
\usepackage{subcaption}
\usepackage{float}
\usepackage{soul}
\usepackage{url}
\usepackage{microtype} % Improved Tracking and Kerning
\usepackage{mathptmx}
\usepackage{graphicx}
\usepackage{times}
\usepackage{enumitem}
\definecolor{mypink1}{rgb}{0.858, 0.188, 0.478}
\definecolor{mygreen1}{rgb}{0.158, 0.688, 0.158}
\definecolor{mred}{rgb}{.80,.12,.30}

\usepackage{tabu}% http://ctan.org/pkg/tabu

\usepackage[normalem]{ulem}
\definecolor{grey}{rgb}{0.5,0.5,0.5}
\newif\ifnotes
\let\origcite\cite
\renewcommand{\cite}[1]{\ifnotes\mbox{\origcite{#1}}\else \origcite{#1}\fi}

\newcommand{\strikeg}[1]{\ifnotes{\color{grey}{\texorpdfstring{\sout{#1}}{#1}}}\fi}
\newcommand{\add}[1]{\ifnotes{\leavevmode\color{mygreen1}{#1}}\else{#1}\fi}
\newcommand{\replace}[2]{\ifnotes{\strikeg{#1}\add{#2}}\else{#2}\fi}

\newcommand{\sys}{REMAP } %note the space at the end
\newcommand{\syss}{REMAP} %note no space

\usepackage{todonotes} 
\usepackage{xargs}  
\definecolor{bubblegum}{rgb}{0.99, 0.76, 0.8}
\newcommandx{\adam}[2][1=]{\todo[linecolor=bubblegum,backgroundcolor=bubblegum!25,bordercolor=bubblegum,inline,#1]{Adam: #2}}

\onlineid{1135}

\vgtccategory{Research}
\vgtcpapertype{system}

\title{Ablate, Variate, and Contemplate: \\ Visual Analytics for Discovering Neural Architectures}

\author{Dylan Cashman, Adam Perer, Remco Chang, Hendrik Strobelt}
\authorfooter{
\item
 Dylan Cashman and Remco Chang are with Tufts University, USA E-mail: dcashm01@cs.tufts.edu, remco@cs.tufts.edu
\item
 Adam Perer is with Carnegie Mellon University, USA E-mail: adamperer@cmu.edu
\item
 Hendrik Strobelt is with the MIT IBM Watson AI Lab, USA E-mail: hendrik.strobelt@ibm.com
}

\shortauthortitle{Cashman \MakeLowercase{\textit{et al.}}: Ablate, Variate, and Contemplate}
\abstract{ %PLACEHOLDER --
Deep learning models require the configuration of many layers and parameters in order to get good results. 
However, there are currently few systematic guidelines for how to configure a successful model.
This means model builders often have to experiment with different configurations by manually programming different architectures (which is tedious and time consuming) or rely on purely automated approaches to generate and train the architectures (which is expensive).
In this paper, we present Rapid Exploration of Model Architectures and Parameters, or \syss, a visual analytics tool that allows a model builder to discover a deep learning model quickly via exploration and rapid experimentation of neural network architectures.
In \syss, the user explores the large and complex parameter space for neural network architectures using a combination of global inspection and local experimentation.
Through a visual overview of a set of models, the user identifies interesting clusters of architectures.
Based on their findings, the user can run ablation and variation experiments to identify the effects of adding, removing, or replacing layers in a given architecture and generate new models accordingly.  They can also handcraft new models using a simple graphical interface.
As a result, a model builder can build deep learning models quickly, efficiently, and without manual programming.
We inform the design of \sys through a design study with four deep learning model builders.  Through a use case\strikeg{ and validation study}, we demonstrate that \sys allows users to discover performant neural network architectures efficiently using visual exploration and user-defined semi-automated searches through the model space.
} % end of abstract


\CCScatlist{ % not used in journal version
 \CCScat{K.6.1}{Management of Computing and Information Systems}%
{Project and People Management}{Life Cycle};
 \CCScat{K.7.m}{The Computing Profession}{Miscellaneous}{Ethics}
}

\teaser{
    \centering
    \includegraphics[width=1\linewidth]{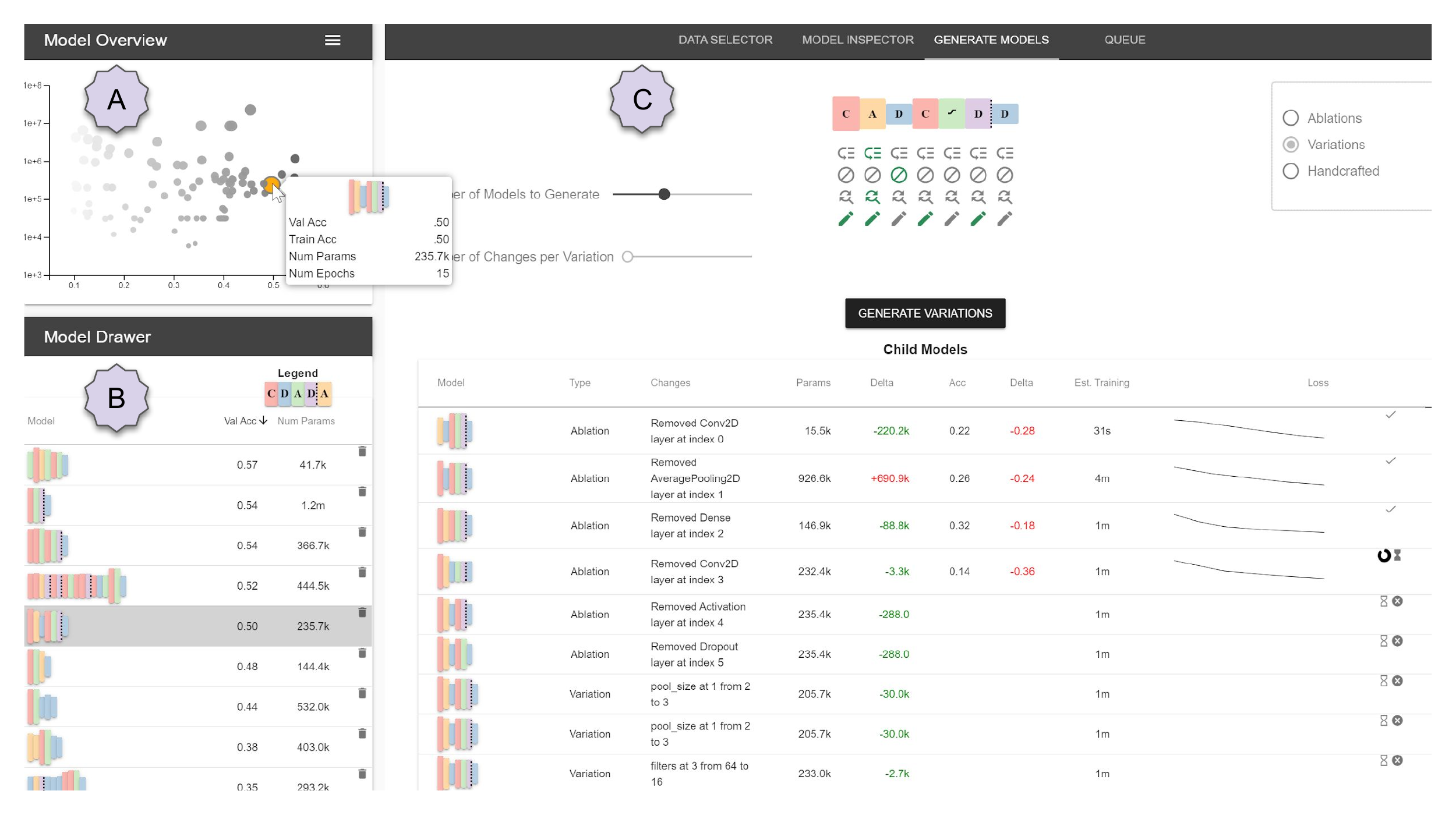}
    \caption{A screenshot of the \sys system.  In the model overview, section A, a visual overview of the set of sampled models is shown. Darkness of circles encodes performance of the models, and radius encodes the number of parameters.  In the model drawer, section B, users can save models during their exploration for comparison or to return to later.  In section C, four tabs help the user explore the model space and generate new models.  The Generate Models tab, currently selected, allows for users to create new models via ablations, variations, or handcrafted templates. }
    \label{fig:interface_overview}
}

\begin{document}


\firstsection{Introduction}

\maketitle
% \dylan{NOTES FOR FINAL DAY - really need to push that there is a tradeoff between model size and performance, and VA systems are good for that.  Look at the conclusions of the use case.}

% \dylan{Source for figures is available at https://docs.google.com/presentation/d/1AydruakiLnqXFUSm7y5dfxE5BdbFZDRDRzcYUXOoQDg/edit?usp=sharing}

% \dylan{Make the point that this can be used for small models, which can be very useful - fast to evaluate, live on edge devices, phones, tablets.  This gets us around computational scaling (because bigger models take much too long to train in situ), but also around visual scaling (we can't really encode RESNET 100)}

Deep neural networks have been applied very successfully in recent advances in computer vision, natural language processing, machine translation and many other domains.  However, in order to obtain good performance, model developers must configure many layers and parameters carefully.  Issues with such manual configuration have been raised as early as 1989, where Miller et al.\cite{miller1989designing} suggested automated neural architecture search should be useful in enabling a wider audience to use neural networks: 

\begin{quotation}``Designing neural networks is hard for humans.  Even small networks can behave in ways that defy comprehension; large, multi-layer, nonlinear networks can be downright mystifying.''\cite{miller1989designing}\end{quotation} 

However, thirty years later, the authors' note is still a common refrain.  While research has continued in automated neural architecture search, much of the progress in algorithms has focused on developing more performant models using prohibitively expensive resources, e.g., state of the art algorithms in reinforcement learning taking 1800 GPU days\cite{DBLP:journals/corr/ZophVSL17} and evolutionary algorithms taking 3150 GPU days\cite{DBLP:journals/corr/abs-1802-01548} to discover their reported architectures. Those users that have access to the type of hardware necessary to use these algorithms likely would either have the expertise needed to manually construct their own network or would have access to a machine learning expert that would be able to do it for them.  

Likewise, a number of visual analytics tools have been released that make neural networks more interpretable and customizable\cite{hohman2018visual}.  However, these tools presuppose that a sufficiently performant model architecture has been chosen a priori without the aid of a visual analytics tool.  The initial choice of neural network architecture is still a significant barrier to access that limits the usability of neural networks.  Tools are needed to provide a human-driven search for neural network architectures to provide a data scientist with an initial performant model.  Once this model has been found, existing visual analytics tools could be used to fine tune it, if needed.

In this work, we present \syss, a tool for human in the loop neural architecture search.  Compared to the manual discovery of neural architectures (which is tedious and time consuming), \sys allows a model builder to discover a deep learning model quickly via exploration and rapid experimentation.  In contrast to fully automated algorithms for architecture search (which are expensive and difficult to control), \sys uses a semi-automated approach where users have fine-grained control over the types of models that are generated.  This allows users to trade off between the size of the model, the performance on individual classes, and the overall performance of the resulting model. 

Through a set of interviews with model builders, we establish a set of tasks used in the manual discovery of neural network architectures.  After developing an initial version of \syss, we held a validation study with the same experts and incorporated their feedback into the tool.  In \syss, users first explore an overview of a set of pre-trained small models to find interesting clusters of models.  \replace{Then, users can explore new models, generated either through small, local architecture searches of ablations and variations of selected models.}{Then, users guide the discovery of new models via two operations on existing models: ablations, in which a new model is generated by removing a single layer of an existing model, and variations, in which several new models are generated by random atomic changes of an existing model, such as a reparameterization or the replacement of an existing layer.}  Users can also manually construct or modify any architecture via a simple drag-and-drop interface.  By enabling global and local inspection of networks and allowing for user-directed exploration of the model space, \sys supports model selection of neural network architectures for data scientists. 

% \textit{Model selection} is the task of constructing and choosing the best available machine learning model given a task and a dataset.  The definition of \textit{best} is context-sensitive in both a computational sense (because there may be multiple objectives being maximized concurrently) but also in a practical sense in that the model builder and their team may have insight about the task at hand that is not captured by the learning algorithms being used.  In order to gather context on the performance of models, the model builder must then gain an understanding of the model space --- how a change in one parameter might effect the ensuing model and its predictions.  

\strikeg{Visual analytics applications have proven invaluable for model selection in machine learning because they can encode models into visualizations, such as scatterplots, and facilitate exploration and understanding of a model space.  Such tools have been used to explore model spaces for decision trees\cite{muhlbacher2018treepod}, regression\cite{muhlbacher2013partition}, clustering\cite{nam2007clustersculptor, cavallo2018clustrophile, kwon2018clustervision, Sacha2018}, classification\cite{van2011baobabview, choo2010ivisclassifier, Krause:2016}, dimensionality reduction\cite{choo2013interactive, jeong2009ipca, nam2013tripadvisor, anand2012visual, liu2015visual}, and time series analysis\cite{bogl2013visual}.}

% In this work, we present a tool for the exploration of model space for feed-forward neural networks.  Neural networks are...

The model space for neural networks poses unique challenges for our tool.  Whereas many of the parameter spaces explored in other types of models have a set number of choices of parameters, the parameter space for neural networks is potentially infinite - one can always choose to add more layers to a network. \strikeg{This makes choosing a computational representation for the model space difficult.  For example, a distance metric is necessary to project all models into a scatter plot in order to view clusters of models.  The goal of \sys is to help the user develop an understanding of the model space, w} In order to aid in the interpretation of the model space, we propose 2-D projections based on two different distance metrics \add{for embedding neural networks} based on Lipton's two forms of model interpretability, \textit{transparency} and \textit{post-hoc interpretability}\cite{lipton2016mythos}.  \strikeg{For the former, we use OTMANN, an optimal-transport based distance, used in the automated machine learning literature, that represents structural similarity between models\cite{kandasamy2018neural}.  For post-hocinterpretability, we use a distance metric that places models close together if their predictions are similar.  By navigating these two spaces, users can get a sense of how structural changes in the model space can affect the predictive qualities of a model.}

The second significant hurdle for a visual model selection over neural networks is to find a visual encoding for neural networks that enabled comparison of many networks while still conveying shape and computation of those networks.  In this work, we contribute a novel visual encoding, called Sequential Neural Architecture Chips (SNACs), which are a space-efficient, adaptable encoding for feed-forward neural networks. SNACs can be incorporated into both visual analytics systems and static documents such as academic papers and industry white papers.  

The workflow of our system largely follows the conceptual framework for visual parameter space analysis from Sedlmair et. al.\cite{sedlmair2014}.  A starting set of models is initially sampled from the space in a preprocessing stage, and projections of the models are calculated.  Models are then explored in three derived spaces: two MDS projections corresponding to the two distance metrics as well as a third projection with interpretable axes.  The system then uses the \textit{global-to-local} strategy of navigating the parameter space, moving from an overview of models to an inspection of individual models in neighborhoods in the derived spaces.  During exploration, users can instruct the system to spawn additional models in the neighborhood of already-sampled models, rendering more definition in their mental model of the parameter space on the regions they are most interested in.

% To inform the development of our system, we conducted a design study with four model builders.  Prior to the development of the tool, participants were asked to describe their process for finding and fine tuning neural networks.  They also offered suggestions for what pain points could be assuaged by a visual analytics system.  From these interviews, we identified a set of tasks in neural architecture search that were used in the manual discovery of neural network architectures.  Several months later, the same four experts used the system to search for models on a well known image classification dataset, and provided qualitative feedback on the system.  This second round of interviews resulted in an additional set of features for fine tuning models.

Overall, the contributions of this paper include:
    \begin{itemize}[itemsep=0pt, topsep=3pt, partopsep=3pt, leftmargin=9pt]

        \item \syss, a visual analytics system for semi-automated neural architecture search that is more efficient than existing manual or fully-automated approaches
        \item A set of visual encodings and embedding techniques for visualizing and comparing a large number of sequential neural network architectures
        \item A set of design goals derived from a design study with four model builders 
        \item A use case applying \sys to discover convolutional neural networks for classification of sketches
                % \adam{I updated this from a design goal to requirements, as Remco suggested.}
    \end{itemize}
    
% \dylan{the contributions should be reordered - the user study is the primary contribution according to the abstract.  Also, the results of the study should be summarized in the bullet corresponding to the study.}

\section{Motivation}

\label{sec:network_selection}
% \subsection{Neural Networks}

% \dylan{ We can also put in a figure here showing a neural network, for explanation's sake and also as a favorable comparison for when we show the SNACs.  Also, we want to exactly delineate which hyperparameters we are searching over - we aren't doing the parameters relating to training, like the learning rate, the batch size, etc. because those aren't related to the architecture.  Point out that these are all orthogonal choices to the architecture.  Once an architecture is chosen, these can be tried.  Look up a citation if there is any automated training program that tries all of these.}

A machine learning \textit{model} is an algorithm that predicts a target label from a set of predictor variables.  These models learn how to make their prediction by learning the relationships between the predictor variables and target label on a \textit{training dataset}.  Machine learning models typically train by iterating over the training set multiple times; each iteration is called an \textit{epoch}.  In each epoch, the model makes predictions and accrues \textit{loss} when it makes poor predictions.  It then updates its learned parameters based on that loss.  At each epoch, the accuracy of the model on a held out portion of the dataset, called the \textit{validation dataset}, is calculated.  

Neural networks are a class of machine learning models that are inspired by the message passing mechanisms found between neurons in brains.  A neural network consists of an architecture and corresponding parameters\footnote{Parameters chosen by the model builder are sometimes called hyperparameters to differentiate from the parameters of a model that are learned during training.  In this work, we call both of these terms parameters, but refer to the latter as learned parameters for the sake of delineation.} chosen by the model builder for each component of that architecture.  The architecture defines the computational graph mapping from input to output, e.g. how the input space, such as an image, is transformed into the output space, such as a classification (the image is a \textit{cat} or a \textit{dog}).  In sequential neural networks, which have simple computation graphs representable by linked lists, the nodes of the computations graphs are called \textit{layers}.  

\strikeg{Different problem domains call for different types of architectures; while image classification typically makes use of sequential neural networks, in which layers are stacked between the input and output, natural language processing networks typically use recurrent neural networks, which take into account temporal effects in their architecture.  In this work, we consider choosing a model for image classification for the sake of scoping the problem down.  This consists of selecting a sequence of layers that feed into one another, as well as the parameters for each layer.}

% \dylan{need one more sentence here explaining that layers have types and each type has its own parameters to choose}

Choosing an architecture that performs well can be difficult\cite{miller1989designing}.  Small changes in parameters chosen by model builders can result in large changes in performance, and many configurations will result in models that quickly plateau without gaining much predictive capacity through training.  In addition, training neural networks is very slow relative to other machine learning methods\add{.  As a result, the process of manually discovering a performant model can be frustrating and costly in time and resources}.  \strikeg{It may be many minutes, hours, or even days before a network has spent enough time training to know if it is a good architecture or not.  Since each layer of a network has its own set of parameters that must be chosen, and the performance of a layer is dependent on all other layers in the network, the space of models is exponentially large.  As a result, a model builder must train many architectures to explore the space and find an architecture that is performant.  Model builders may also be concerned with the size of the model if the model needs to be deployed on mobile or internet of things devices.  The connection between adding or removing layers can have an unexpected effect on the size of a model; adding a fully connected layer might result in millions more parameters or just thousands, depending on if it is added early or late in the architecture.}

% \dylan{Any thoughts on explaining why "small" is good here?  I feel like we're being redundant with the intro here.}

Automated algorithms for neural architecture search generate thousands of architectures in order to find performant architectures\cite{zoph2016neural} and can require tens of thousands of GPU hours of training\cite{DBLP:journals/corr/ZophVSL17, DBLP:journals/corr/abs-1802-01548}.  The best discovered models might be too large for a model builder if they aim to deploy their model on an edge device such as a tablet or an internet of things device.  Ideally, they would be able to handcraft each generated model and monitor its training to not waste time and resources discovering models that were not useful.  However, handcrafting each model can be time consuming and repetitive.

In our tool, we seek a middle ground.  We initially sample a small set of architectures, and then use visualizations to facilitate exploration of the model space.  Model builders can find regions of the space that produce models they are interested in, and then they can execute a local, constrained, automated search near those models.  As they get closer to finding an acceptable model, they can explicitly handcraft models through a graphical interface.  Rather than training thousands of architectures, the model builder trains orders of magnitude less, and stops the architecture search when they have found an acceptable model.  Our semi-automated approach lets the user search for neural architectures without the tedium of manually constructing each model and without the resources and time required by fully-automated algorithms for neural architecture search.

% We note that there are a number of training parameters that are experimented with to maximize performance of a network, such as learning rate schedulers and data augmentation strategies, that we do not include in our model search.  If needed, a model builder could experiment with these parameters after using our tool to find a performant architecture.

% In this paper, we demonstrate our techniques on sequential neural architectures for image classification.  We believe this is representative for a large class of commonly used deep neural networks.  
% For example, when building a decision tree, there are a smaller set of hyperparameters that must be chosen, such as the max depth of the tree, the pruning strategy, and the splitting strategy.  In contrast, to 
% To build a neural network, a \hen{model architect} must first choose the first layer and its corresponding hyper parameters, then choose the next layer, and eventually decide when to stop adding layers.  Even if the number of layers is constrained to $n$ layers, the number of potential models is exponential in $n$ without taking into account that each layer has its own space of continuous parameters.

% \remco{if this section is called motivation, there should be another paragraph here to summarize what the paper aims to accomplish. In this sense, I might suggest a re-organization. Current the flow is: background info, what we do, challenges. I would swap it around so that it sounds more like: background info, challenges, what we do.} \adam{+1} \hen{+2}

\section{Related Work}

\subsection{Neural Architecture Search}
Algorithms for the automated discovery of neural network architectures were proposed as early as the late 1980s using genetic algorithms\cite{miller1989designing}.  Algorithm designers were concerned that neural networks were excessively hard to implement due to their large parameter space and odd reaction to poor parameterizations.  In recent years, interest in neural networks has exploded as they have proven to be state of the art algorithms for image classification\cite{krizhevsky2012imagenet}, text classification\cite{lai2015recurrent}, video classification\cite{karpathy2014large}, image captioning\cite{xu2015show}, visual question answering\cite{lu2016hierarchical}, and a host of other classic artificial intelligence problems.  An increased interest in automated neural architecture searches has followed, resulting in a variety of algorithms using Bayesian optimization\cite{snoek2012practical}, network morphisms\cite{jin2018efficient}, or reinforcement learning\cite{zoph2016neural, DBLP:journals/corr/BakerGNR16}.  These algorithms typically define the architecture space so that it is easily searchable by classical parameter space exploration techniques, such as gradient-based optimization\cite{kandasamy2018neural, DBLP:journals/corr/abs-1806-09055}.  Elsken et al. provide a summary of new research in algorithmic methods in a recent survey\cite{elsken2018neural}.  

Such methods are driven by an attempt to compete with state of the art performant architectures such as ResNet\cite{he2016deep} or VGGNet\cite{Simonyan14c} that were carefully handcrafted based on years of incremental research in the community.  Because performance has been the primary motivator, automated neural architecture search algorithm designers have depended on expensive hardware setups using multiple expensive GPUs and very long search and training times\cite{DBLP:journals/corr/abs-1806-09055}.  As a result, the use of these algorithms is out of reach for many potential users without expensive hardware purchases or large outlays to cloud machine learning services.  
In contrast, our tool is more accessible to data scientists because it drastically shrinks the search space by conducting user-defined local, constrained searches in neighborhoods around models the user is interested in.  \strikeg{These searches also skip over any models that are above a certain size threshold.  By designing our search around generating relatively small models and only searching in local, interesting spaces, we can have a less resource-intensive search.  This makes \sys able to reach a new set of users compared to fully automated algorithms.}

% Neural Network interpretability and accessibility
\subsection{Visualization for Neural Networks}

Visualization has been used in both the machine learning literature and the visual analytics literature for understanding and diagnosing neural networks.  In particular, attempts have been made to explain the decision making process of trained networks.  Saliency maps\cite{DBLP:journals/corr/SimonyanVZ13} and gradient-based methods\cite{DBLP:journals/corr/SelvarajuDVCPB16} were an early attempt to understand which pixels were most salient to a network's predictions in image classification networks.  However, recent work has shown that saliency maps may be dependent only on inherent aspects of the image and not the network's decision making, calling into doubt some of the truthfulness of such methods\cite{DBLP:journals/corr/abs-1810-03292}.  Methods also exist which inspect the effect of individual layers on the decisions of the network\cite{zeiler2014visualizing, DBLP:journals/corr/YosinskiCNFL15}.  \textit{Lucid} is a library built on the \textit{Tensorflow} machine learning library for generating various visualizations of networks\cite{olah2018the}. 
% These methods seek to explain the decision making of a network after it has been chosen and trained.  They offer some insight on the importance of various layers and components of the architecture, but typically are used 

Visual analytics tools extend these techniques by offering interactive environments for users to explore their networks.  \replace{CNNVis allows users to inspect bundles of neurons throughout a network to understand the different decision making units that contribute to the ultimate prediction\cite{liu2017towards}.  LSTMVis similarly shows the user subsets of neurons whose activations correlate with aspects of the model's inference, although in a recurrent neural network rather than a convolutional neural network\cite{strobelt2018lstmvis}. DQNVis aims to explain how Q-networks are able to perform so well at  complex artificial intelligence tasks such as playing Atari games by extracing and visualizing action/reward patterns\cite{wang2019dqnviz}.  Industry-scale networks can be difficult to visualize due to the number of layers used; interactive tools can prove useful for exploring such networks\cite{kahng2018cti, wongsuphasawat2018visualizing}.  Tensorflow Playground\cite{tfplayground} and GAN Lab\cite{kahng2019gan} are two instructional tools, allowing users to interactively control and build small networks to understand the influence of various hyperparameter choices.}{Some tools allow users to inspect how various components of a trained network contribute to its predictions\cite{liu2017towards, strobelt2018lstmvis, wang2019dqnviz, kahng2018cti, wongsuphasawat2018visualizing}, while others allow the user to build and train toy models to understand the influence of various hyperparameter choices\cite{tfplayground, kahng2019gan}}
Other tools focus on debugging a network to determine which changes must be made to improve its performance\replace{.  RNNbow is a tool for visualizing gradient flow during training that helps indicate if a recurrent neural network is not learning long term dependencies\cite{cashman2017rnnbow}.  Seq-2-Seq-Vis visualizes all five components of the Seq-2-Seq model for language translation, helping the user identify which component is responsible for particular errors in inference\cite{strobelt2019s}.  DGMTracker diagnoses deep generative models by visualizing neurons contributing to failure during training\cite{liu2018analyzing}.  DeepEyes exposes errors during the training process by showing the user activation patterns and degenerate filters so that the user can intervene during an errant training process\cite{pezzotti2017deepeyes}.}{ by viewing the activations, gradients, and failure cases of the network\cite{cashman2017rnnbow, strobelt2019s, liu2018analyzing, pezzotti2017deepeyes}.}  Hohman et al. provide a comprehensive overview of visual analytics for deep learning\cite{hohman2018visual} .  

All of these visual analytics tools presuppose that the user has selected an architecture and wants to inspect, explain, or diagnose it.  In contrast, \sys allows the user to discover a new architecture.  A user of \sys might take the discovered architecture and then feed it into a tool such as DeepEyes to more acutely fine tune it for maximal performance\cite{pezzotti2017deepeyes}.

% Pezzotti (fine tuning and diagnosing a chosen network)
% LSTMVis, Seq2SeqVis (fine tuning and diagnosing a chosen recurrent network for text)
% Shixia's students' works on inspecting / diagnosing networks
% DeepVis (Minsuhk)
% Tensorflow vis / tensorboard

\subsection{Visual Analytics for Model Selection}

Model selection is highly dependent on the needs of the user and the deployment scenario of a model.  Interactivity can be helpful in comparing multiple models and their predictions on a holdout set of data. Zhang et. al. recently developed Manifold, a framework for interpreting machine learning models that allowed for pairwise comparisons of various models on the same validation data\cite{zhang2018manifold}. M\"uhlbacher and Piringer support analyzing and comparing regression models based on visualization of feature dependencies and model residuals\cite{muhlbacher2013partition}.  Schneider et al. demonstrate how the visual integration of the data and the model space can help users select relevant classifiers to form an ensemble\cite{schneider2018integrating}.  Snowcat is a visual analytics tool that enables model selection from a set of black box models returned from a automated machine learning backend by visually comparing their predictions in the context of the data source\cite{cashman2018visual}.  These methods all assume that the model is being selected from a set of pretrained models, in contrast to our system which can generate additional models based on user input.

% Sedlmair et al. analyzed the visualization literature to develop a framework for visual parameter space analysis, identifying common components of such tools\cite{sedlmair2014}.  \sys follows a common workflow identified in that work by initially sampling the parameter space, then allowing global inspection of those models before moving to local trial and error in the neighborhood of interesting models.

\add{
\subsection{Visual Analytics for autoML}
Automated Machine Learning, or autoML, comprises a set of techniques designed to automate the end-to-end process of ML.
To accomplish this, autoML techniques automate a range of ML operations, including but not limited to, data cleaning, data pre-processing, feature engineering, feature selection, algorithm selection and hyperparameter optimization\cite{guyon2015design}.  Different autoML libraries such as AutoWeka\cite{thornton2013auto, kotthoff2016auto}, Hyperopt\cite{bergstra2013hyperopt, komer2014hyperopt}, and Google Cloud AutoML\cite{googlecloudautoml} are in use either commercially or as open source tools. 

Visual Analytics systems have been used to both provide an interface to the autoML process as well as insert a human in the loop to various parts of the process.  }
\strikeg{\sys is most similar to two tools that let the user search through an initial set of models and generate new children of those models based on user interactions.}  TreePOD\cite{muhlbacher2018treepod} helps users balance potentially conflicting objectives such as accuracy and interpretability of automatically generated decision tree models by facilitating comparison of candidate tree models.  Users can then spawn similar decision trees by providing variation parameters, such as tree depth and rule inclusion.  BEAMES\cite{dasbeames} allows users to search for regression models by offering feedback on an initial set of models and their predictions on a held out validation dataset.  The system spawns new models based on that feedback, and users iterate until they find a satisfactory model.\strikeg{\sys similarly takes in iterative feedback on where to spawn new models in the model space.  It uses visual encodings of neural network architectures to facilitate visual comparison, compared to the decision trees and squarified tree maps found in TreePOD.}  \add{Various tools facilitate user control over the generation of models for regression\cite{muhlbacher2013partition}, clustering\cite{nam2007clustersculptor, cavallo2018clustrophile, kwon2018clustervision, Sacha2018}, classification\cite{van2011baobabview, choo2010ivisclassifier}, dimension reduction\cite{choo2013interactive, jeong2009ipca, nam2013tripadvisor, anand2012visual, liu2015visual}.  \sys differs from those tools in that it explicitly uses properties of neural networks, such as the sequence of layers, in its visual encodings.  Also, because neural networks take much longer to train than decision trees, regression models, and most models considered by previous visual analytics tools, \sys places more of an emphasis on only generating models that the user is interested in.}  \strikeg{Neural networks take much longer to train, and so \sys has to be more sparing in when it spawns a new model, and thus must provide more control in which models to spawn.}

\section{Design Study}
\label{sec:design_study}

% Automated neural architecture searches are able to find effective architectures, but they require resources outside of the norm for a non-machine learning expert.  Instead, we sought to develop a visual analytics tool that would empower data scientists in machine learning to discover performant neural network architectures.  
In order to develop a set of task requirements\strikeg{ for visual analytics of automated neural architecture search}, we interviewed a set of model architects about their practices in manually searching for neural network architectures.  We also asked the experts what visualizations might be helpful for non-experts in a human-in-the-loop system for neural network architecture search.

% \adam{I took a stab are re-doing the below paragraph.  As written, it didn't seem mature enough to be taken seriously.}
\textbf{Participants:} To gather participants, we recruited individuals with experience in designing deep neural network architectures.  Four experienced model builders agreed to participate in the interview study.  Three of the participants are PhD students in machine learning, and the fourth participant has a Masters degree in Computational Data Science and works in industry.  They had previously used neural networks for medical image classification, image segmentation, natural language processing, and graph inference.  One participant contributed to an open source automated neural architecture search library.  All four participants were from different universities or companies and had no role in this project.  Participants were compensated with a twenty dollar gift card.

\textbf{Method:}  
Interviews were held with each participant to establish a set of tasks used to manually discover and tune neural networks.  The interviews were held one-on-one using an online conferencing software with an author of this work and took one hour each.  Audio was recorded and transcribed with the participants' consents so that quotes could be taken.

Interviews were semi-structured, with each participant being asked the same set of open-ended questions\footnote{Interview questions are available as a supplemental document.}.  They were first asked to describe their work with neural networks, including what types of data they had worked with.  They were then asked about their typical workflow in choosing and fine tuning a model.  Then, the benefits of human-in-the-loop systems for neural network model selection were discussed.  Lastly, participants were prompted for what types of features might be useful in a visual analytics system for selecting a neural network.

% \remco{this part is a little weird in that you don't show what the initial version of the system looks like. As a result, a reader doesn't learn much of anything from your effort of doing this study as iterative design. If space permits, I would expand on this similar to the way that we discussed Snowcat. Along this line, I would break up this section into two sections -- one for each round of interview. This way your writing doesn't ping-poing from talking about the two rounds of interviews, then back to talking about the finding from the first.}

\textbf{Findings:}
The findings from the interview study resulted in the following set of design goals. 

%  In this setting, this means removing one layer at a time to see the influence of each layer on the architecture's performance.  
\begin{itemize}[itemsep=0pt, topsep=3pt, partopsep=3pt, leftmargin=9pt]
    \item \textbf{Goal G1: Find Baseline Model:} Three out of the four participants noted that when they are building an architecture for a new dataset, they start with a network that they know is performant.  This network might be from a previous work in the literature or it might be a network they've used for a different dataset.  This network typically provides a baseline, upon which they then do fine tuning experiments: "\textit{The first step is just use a structure proposed in the paper.  Second step I always do is to change hyperparameters.  For example, I add another layer or use different dropouts.}"  One participant noted that they prioritize using a small model as a baseline because they are more confident in the stability of small models, and it is easier to run fine tuning experiments on small models because they train faster.
    \item \textbf{Goal G2: Generate Ablations and Variations:}  Three participants noted that in order to drive their fine tuning, they typically do two types of experiments on a performant network.  First, they do ablation studies, a technical term referring to a set of controlled experiments in which one independent variable is turned off for each run of the experiment. Based on the results of the ablation studies, they then generate variations of the architecture by switching out or reparameterizing layers that were shown to be less useful by the ablations.  Two participants noted that these studies can be onerous to run, since they need to write code for each version of the architecture they try.
    \item \textbf{Goal G3: Explain/Understand Architectures:}  When asked about the types of information to visualize for data scientists, two participants noted that \replace{visualizations of the architecture can help a user to glean}{users might be able to glean} a better understanding of how neural networks are constructed\add{ by viewing the generated architectures}.  While it may be obvious to the study participants that convolutional layers early in the network are good at extracting features but less helpful in later layers, that understanding comes from experience.  By visually comparing models, non-experts might come to similar conclusions. One participant pointed out that the human in the loop could interpret the resulting model more, helping "\textit{two people, the person developing the results, and the person buying the algorithm}."  
    \item \textbf{Goal G4: Human-supplied Constrained Search:}  Participants were asked what role a human in the loop would have in selecting a neural network architecture, compared to a fully-automated model search.  All four participants noted that if the data is clean and correctly labeled, and there are sufficient resources and time, that a human in the loop would not improve upon an automated neural architecture search.  But three participants noted that when resources are limited, the human user can compensate by offering constraints to an automated search, pointing an automated search to particular parts of the model space that are more interesting to the user.  One participant noted that for fully-automated model search, "\textit{some use reinforcement learning, [some] use Bayesian optimization.  The human can also be the controller.}"  
\end{itemize}
% DARPA's Data-Driven Discovery of Models project

From these findings, we distill the following tasks that our system must support to enable data scientists to discover performant neural network architectures.  

\begin{figure*}[thb]
\centering
\subcaptionbox{\label{fig:model_inspection}}{\includegraphics[width=0.49\textwidth, height=2.0in]{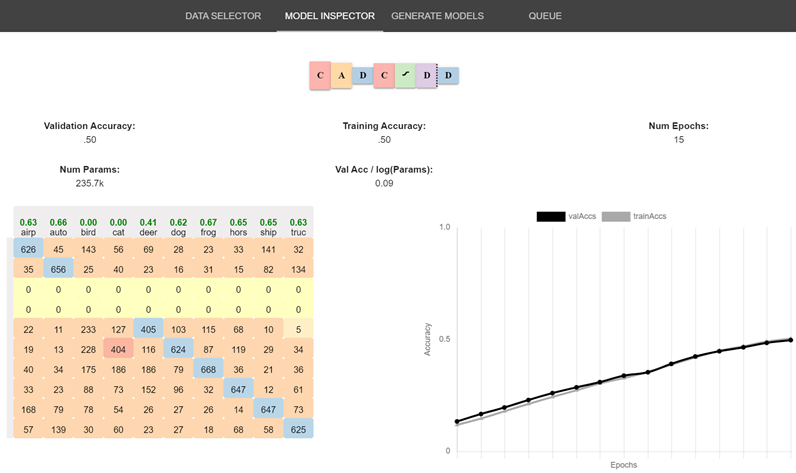}}%
%\bigskip
\quad%\hfill
%\par\bigskip
%\vline    \hspace{0.1cm}
\subcaptionbox{\label{fig:data_selector}}{\includegraphics[width=0.49\textwidth, height=2.0in]{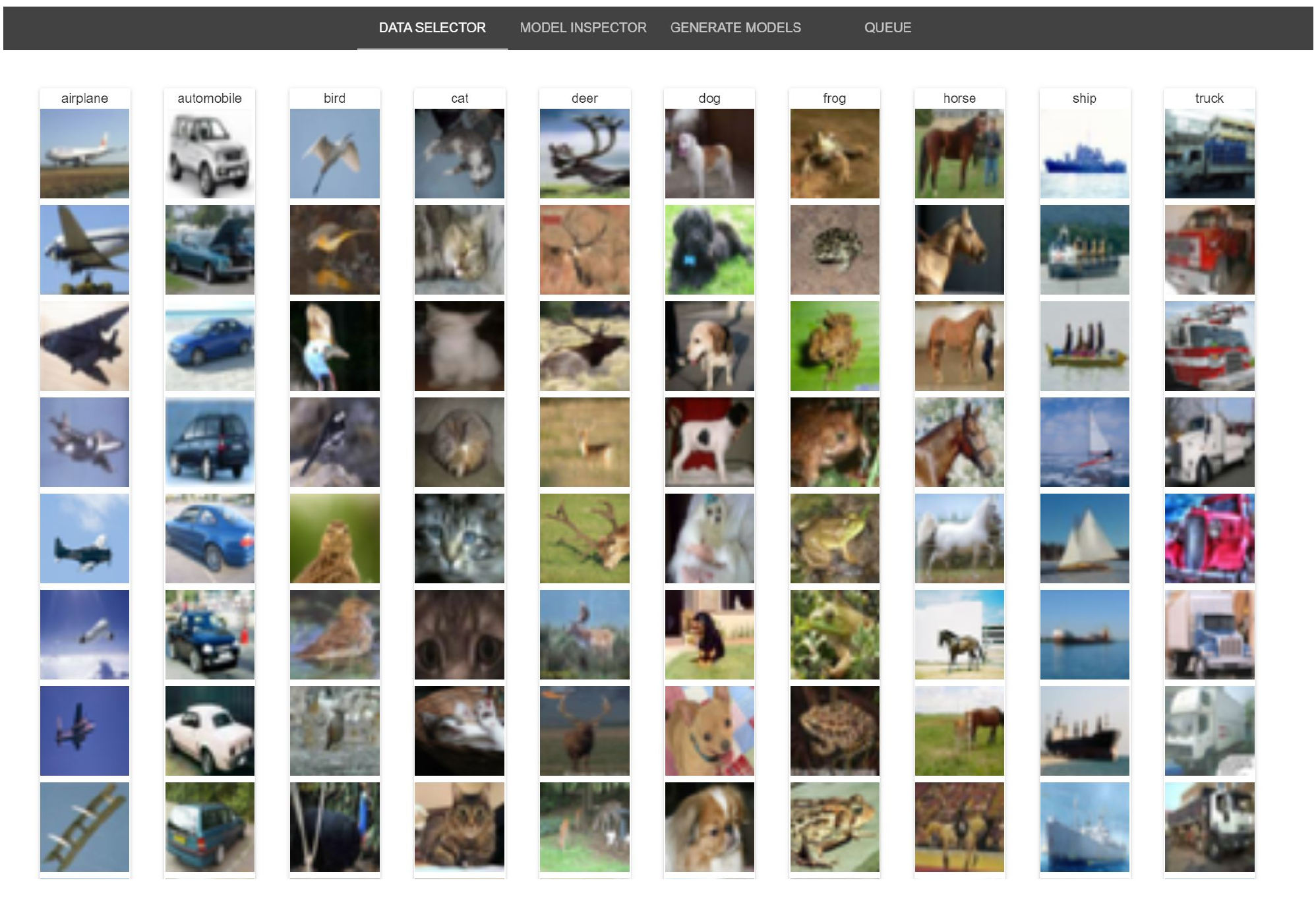}}%
\caption{(a) The model inspection tab lets users see more granular information about a highlighted model.  This includes a confusion matrix showing which classes the model performs best on or misclassifies most frequently.  Users can also view training curves to determine if an architecture might be able to continue to improve if trained further. (b) By selecting individual classes from the validation data, users can update the darkness of circles in the the model overview to see how all models perform on a given class.}
\end{figure*}

\begin{itemize}[itemsep=0pt, topsep=3pt, partopsep=3pt, leftmargin=9pt]
\item \textbf{Task T1: Quickly search for baseline architectures through an overview of models.} Users must be able to start from an effective baseline architecture [G1].  Experts typically refer to the literature to find a starting architecture that has already been shown to work on a similar problem, such as VGGNet\cite{Simonyan14c} or ResNet\cite{he2016deep}.  These models, however, have hundreds of millions of parameters and cannot be easily and quickly experimented upon, so some other manner for finding compact, easily trainable baseline models is needed.  Users should be able to find small, performant baseline models easily via visual exploration.  

\item \textbf{Task T2: Generate local, constrained searches in the neighborhood of baseline models.} Our tool needs to provide the ability to explore and experiment on baseline models using ablations and variations [G2].  These experiments should help the user in identifying superfluous layers in an architecture.  The human user should be able to provide simple constraints to the search for new architectures [G4].  

\item \textbf{Task T3: Visually compare subsets of models to understand small, local differences in architecture. } The tool should support visual comparisons of models to help the user understand what components make a successful neural network architecture.  This helps the user interpret the discovered neural network models [G3] while also informing the user's strategies for generating variations and exploring the model space [G4].

\end{itemize}

Beyond these three tasks, we also note that compared to many fully automated neural architecture searches, we must be cognizant of limitations on resources.  Much of the neural network literature assumes access to prohibitively expensive hardware and expects the user to wait hours or days for a model to train.  In our tool, we focus instead on small models that are trainable on more typical hardware.  While these models may not be state of the art, they are accessible to a much wider audience.

\section{REMAP: Rapid Exploration of Model Architectures and Parameters }

% \begin{figure*}
%     \centering
%     \includegraphics[width=1\linewidth]{figs/adam/adam-fig2.pdf}
%     \caption{A screenshot of the \sys system.  In the model overview, section A, a visual overview of the set of sampled models is shown. Color of circles encodes performance of the models, and radius encodes the number of parameters.  In the model drawer, section B, users can save models during their exploration for comparison or to return to later.  In section C, four tabs help the user explore the model space and generate new models.  The Generate Models tab, currently selected, allows for users to create new models via ablations, variations, or handcrafted templates. }
%     \label{fig:interface_overview}
% \end{figure*}

\sys is a client-server application that enables users to interactively explore and discover neural network architectures.\footnote{The source code for the tool along with installation instructions are publicly available at \url{https://github.com/dylancashman/remap_nas}.}
% It is designed to enable non-experts to be able to do similar tasks for model selection as done manually through programming by machine learning experts.  
A screenshot of the tool can be seen in Figure~\ref{fig:interface_overview}.  The interface features three components: a model overview represented by a scatter plot (Fig.~\ref{fig:interface_overview}A), a model drawer for retaining a subset of interesting models during analysis (Fig.~\ref{fig:interface_overview}B), and a data/model inspection panel (Fig.~\ref{fig:interface_overview}C).  

% All screenshots in this section used the CIFAR-10 dataset, a collection of 50,000 training images and 10,000 testing images each labeled as one of ten mutually exclusive classes\cite{krizhevsky2009learning}.  Each image is colored and 32x32 pixels.  It is a typical image classification dataset used for validation in deep learning research.  

All screenshots in this section use the CIFAR-10 dataset, a collection of 50,000 training images and 10,000 testing images each labeled as one of ten mutually exclusive classes\cite{krizhevsky2009learning}.   Model training including both preprocessing and in-situ model generation was done using a Dell XPS 15 laptop with a 2.2ghz i7-8750 processor, 32 GB of ram, and a NVIDIA GeForce GTX 1050 Ti GPU with 4GB of VRAM.

\subsection{General Workflow}

The user workflow for \sys is inspired by the common workflow identified in the interview study and encompasses tasks T1, T2, and T3 as defined in section~\ref{sec:design_study}.  First, they find a baseline model by visually exploring a set of pre-trained models in the \textbf{model overview} [T1], seen in Figure~\ref{fig:interface_overview}A.  They select models of interest by clicking on their respective circles, placing them into the model drawer, seen in Figure~\ref{fig:interface_overview}B.  By mousing over models in the overview and scanning the model drawer, users can visually compare models of interest [T2].  Then, they use the \textbf{ablation} and \textbf{variation} tools [T3] to fine tune each model of interest, as seen in Figure~\ref{fig:interface_overview}C.  These tools spawn new models with slightly modified architectures that train in the background, which in turn get embedded in the model overview.  Instructions for new models are sent back to the server.  The server maintains a queue of models to train and communicates its status after each epoch of training.

Users iterate between exploring the model space to find interesting baseline models and generating new architectures from those baseline models.  For the types of small models explored in this tool, training can take 1-3 minutes for a single model.  Users can view the current training progress of child models in the Generate Models tab, or can view the history of all training across all models in the \textbf{Queue} tab.  In the queue tab, they can also reorder or cancel models if they don't want to wait for all spawned models to train.

% \begin{figure}
%     \centering
%     %\includegraphics[width=0.4\textwidth]{figs/adam/adam-inspector.pdf}
%     \includegraphics[width=\linewidth]{figs/remap-model-inspector.png}
%     \caption{The model inspection tab lets users see more granular information about a highlighted model.  This includes a confusion matrix showing which classes the model performs best on or misclassifies most frequently.  Users can also view training curves to determine if an architecture might be able to continue to improve if trained further.}
%     \label{fig:model_inspection}
% \end{figure}

% \begin{figure}
%     \centering
%     \includegraphics[width=0.4\textwidth]{figs/adam/adam-selector.pdf}
%     \caption{By selecting individual classes from the validation data, users can change the coloring of models in the the model overview.}
%     \label{fig:data_selector}
% \end{figure}

If users are particularly interested in performance on certain classes in the data, they can select a data class using the data selector seen in Figure~\ref{fig:data_selector} to modify the model overview.  Users can also see a confusion matrix corresponding to each model in the \textbf{model inspection} tab, seen in Figure~\ref{fig:model_inspection}.  By interacting with both the model space and the data space, they are able to find models that match their understanding of the data and the importance of particular classes.

% Note that these steps are common steps in model selection tools as identified in Sedlmair et al.'s work on visual parameter space analysis\cite{sedlmair2014}.  Initially, users are presented with a global view of the parameter space, with some presampled models to inspect.  They move to a local inspection of certain interesting presampled models using three different projections in the model overview described below.  Then, they use the \textit{trial and error} strategy to spawn new models near interesting baselines.  They continue to iterate, checking on new models with the model inspection tool until a satisfactory model is found.

\subsection{Preprocessing}

In order to provide a set of model baselines, \sys must generate a set of initial models.  This set should be diverse in the model space, using many different combinations of layers in order to hopefully cover the space.  That way, whether the user hopes to find a model that performs well on a particular class or that has a particularly small number of parameters, there will exist a reasonable starting point to their model search.

\sys generates this initial model space by using a random scheme based on automated neural architecture searches in the literature\cite{elsken2018neural}.  A Markov Chain is defined which dictates the potential transition probabilities from layer to layer in a newly sampled model.  Starting from an initial state, the first layer is sampled, then its hyperparameters are sampled from a grid.  Then, its succeeding layer is sampled based on what valid transitions are available.  Transition probabilities and layer hyperparameters were chosen based on similar schemes in the autoML literature\cite{DBLP:journals/corr/BakerGNR16}, as well as conventional rules of thumb.  For example, convolutional layers should not follow dense layers because the dense layers remove the locality that convolutional layers depend on.  In essence, \sys uses a small portion of a random automated neural architecture search to initialize the human in the loop search.
% does not need to be thorough in the way that fully automated neural architecture searches are - 
For models in this section and in screenshots, 100 initial models were generated and trained for 10 epochs each, taking approximately 4 hours.  While that is a nontrivial amount of required preprocessing time, it compares favorably to the tens of thousands of GPU hours required by a fully automated search\cite{DBLP:journals/corr/ZophVSL17, DBLP:journals/corr/abs-1802-01548}, which might sample over 10,000 models\cite{zoph2016neural}.

\subsection{Model Overview}

The top left of the interface features the model overview (Fig.~\ref{fig:interface_overview}A), a scatter plot which visualizes three different 2D projections of the set of models.  The user is able to toggle between the different 2D projections.  The visual overview of the model space serves two purposes.  First, it can serve as the starting point for model search, where users can find small, performant baseline models to further analyze and improve.  The default view plots models on interpretable axes of validation accuracy vs. a log scale of the number of parameters, visible in Figure~\ref{fig:interface_overview}.  Each circle represents a trained neural network architecture.  The darkness of the circle encodes the accuracy of the architecture on a held out dataset, with darker circles corresponding to better accuracy.  The radius of the circle encodes the log of the number of parameters.  This means that in the default projection, the validation accuracy and the number of parameters are double encoded - this is based on the finding from the interview study that finding a small, performant baseline model is the first step in model selection.  The lower right edge of the scatter plot forms a Pareto front, where model builders can trade off between performance of a model and its size, similar to the complexity vs. accuracy plots found in Muhlbacher et al.'s TreePOD tool for decision trees\cite{muhlbacher2018treepod}.

% \begin{figure}
%     \centering
%     % \begin{subfigure}
%     %     \centering
%     %     \includegraphics[width=0.45\textwidth]{figs/hamming_proj.png}
%     % \end{subfigure}
%     % \begin{subfigure}
%     %     \centering
%     %     \includegraphics[width=0.45\textwidth]{figs/otmann_proj.png}
%     % \end{subfigure}
%     \includegraphics[width=0.48\textwidth]{figs/adam/adam-overviews.pdf}
%     \caption{Two alternative visual overviews of the model space.  Section (a) shows the set of models on a set of interpretable axes, validation accuracy vs. log of the number of parameters.  Sections (b) and (c)  use multidimensional scaling to layout the same set of models based on prediction similarity (b) and structural similarity (c).  As with Figure~\ref{fig:interface_overview}(a), the darkness of the circle encodes the model accuracy, and the radius of the circle encodes the log of the number of parameters.}
%     \label{fig:projections}
% \end{figure}

Once baseline models have been selected, the model overview can also be used to facilitate comparisons with neighbors of the baseline.  Users are able to view details of neighboring architectures by hovering over their corresponding points in the overview.  By mousing around a neighborhood of an interesting baseline model, they might be able to see how small changes in architecture affect model performance.  However, it is well known that neural networks are notoriously fickle to small changes in parameterization\cite{miller1989designing}.  \replace{As a result, comparing models in the initial view of accuracy vs. number of parameters may not help the user understand the model space, since t}{T}wo points close together in that view could have wildly different architectures.  

To address this, \sys offers two additional projections based on two distance metrics between neural networks.  The two metrics are based on the two types of model interpretability identified in Lipton's recent work\cite{lipton2016mythos}: structural and post-hoc.  Their respective projections are seen in Figure~\ref{fig:projections}, with the same model highlighted in orange in both projections.  2-D Projections are generated from distance metrics using \texttt{scikit-learn}'s implementation of Multidimensional Scaling\cite{pedregosa2011scikit}.

\vspace{3pt}
\noindent \textbf{Structural interpretability} refers to the interpretability of how the components of a model function.  A distance metric based on structural interpretability would place models with similar computational components, or layers, close to each other in the projection.  We used OTMANN distance, an Optimal Transport-based distance metric that measures how difficult it is to transform one network into another, similar to the Wasserstein distance between probability distributions\cite{kandasamy2018neural}.  
% OTMANN distance was developed for an automated neural architecture search as a way of making a Bayesian update on the predicted accuracy of untrained neural networks based on their proximity to trained networks.  
The resulting projection is seen in section B of Figure~\ref{fig:projections}.  Projecting yb this metric allows users to see how similar architectures can result in large variances in validation accuracy and number of parameters.

\vspace{3pt}
\noindent \textbf{Post-hoc interpretability} refers to understanding a model based on its predictions.  A distance metric based on post-hoc interpretability would place models close together in the projection if they have similar predictions on a held-out test set.  Ideally, this notion of similarity should be more sophisticated than simply comparing their accuracy on the entire test set --- it should capture if they usually predict the same even on examples that they classify incorrectly.  We use the edit distance between the two architectures' predictions on the test set.  The resulting projection is seen in section C of Figure~\ref{fig:projections}.  It can be used to find alternative baseline architectures that have similar performance to models of interest.

% In general, machine learning models can live in many different types of spaces.  A visual analytics tool must choose which spaces to show to the user based on the user's goals.  We use three types: the first defined by a set of interpretable axes, the second defined by structural differences, and the third defined by differences in predictions.  To understand the model space, users may need to toggle back and forth between the three spaces, seeing how the neighborhoods of interesting models differ between the three spaces.  
% \dylan{Remove the preceding paragraph?}

New models generated via ablations and variations are embedded in the model overviews via an out-of-sample MDS algorithm\cite{trosset2008out}.  Users can view how spawned models differ from their parent models in the different spaces and get a quick illustration of which qualities were inherited by the parent model.  

\subsection{Ablations and Variations}

\begin{figure}[thb]
\centering
\subcaptionbox{\label{fig:ablation}}{\includegraphics[width=0.35\textwidth]{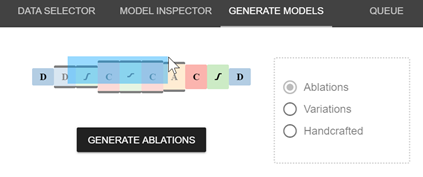}}\\
\subcaptionbox{\label{fig:variations}}{\includegraphics[width=0.35\textwidth]{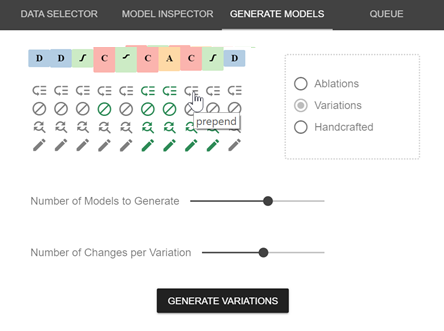}}
\caption{Controls for creating (a) Ablations and (b) Variations.  Users toggle between the two types of model generation with a radio button.  Ablations create a set of models, one for each layer with that layer removed, to communicate the importance of each layer.  The Variations feature runs constrained searches in the neighborhood of a selected model.  Users toggle which types of variations are allowed for each layers, as well as the number of variations allowed per model}
\end{figure}

% \begin{figure}
%     \centering
% %    \begin{subfigure}
% %        \centering
%         \includegraphics[width=0.45\textwidth]{figs/adam/adam-ablations.pdf}
% %    \end{subfigure}
% %    \begin{subfigure}
% %        \centering
%         \includegraphics[width=0.45\textwidth]{figs/adam/adam-variations.pdf}
% %    \end{subfigure}
%     % \includegraphics[width=0.45\textwidth]{figs/two_variations_labeled.png}
    
%     \adam{I tried to update this, but in the google slides, there wasn't an ablation image without a selection.}
%     \caption{Controls for creating Ablations (top (a)) and Variations (bottom (b)).  Users toggle between the two types of model generation with a radio button.  Ablations create a set of models, one for each layer with that layer removed, to communicate the importance of each layer.  Variations are constrained searches in the neighborhood of a selected model.  For variations, users toggle which types of variations are allowed for each layers.  Valid variation types are \textit{prepend} with a new layer, \textit{remove} a layer, \textit{replace a layer}, or \textit{reparameterize} a layer.  Users decide how many variations to generate and how many modifications per variation to allow via slider controls.}
%     \label{fig:variations}
% \end{figure}

According to our expert interviews, an integral task in finding a performant neural network architecture is to run various experiments on slightly modified versions of a baseline architecture.  One type of modification that is done is an ablation study, in which the network is retrained with each feature of interested turned off, one at a time.  The goal of ablations is to determine the effect of each feature of a network.  This might then drive certain features to be pruned, or for those features to be duplicated.

% \dylan{Add ablation image, or put ablations into the variations tab}
% \adam{I recommend using an ablation AND variation image in this section.  I think this is one of the most important sections and you should walk the reader through it with a narrative + figures}

In our system, users can automatically run ablation studies that retrain a selected model without each of its layers. The system will then train those models for the same number of epochs as the parent model, and display to the user the change in validation accuracy.  If the user wants to make a more fine-grained comparison between the models, the user can move the model resulting from an ablation into the model drawer, and then use the model inspector to compare their confusion matrices. 

% Various other fine-tunings are possible; the experts interviewed mentioned that they may reparameterize layers (such as changing an activation function or changing the stride or number of filters in a convolutional layer), or they might switch out or add layers at parts of the network.  
% By exploring the neighborhood of a model of interest in the model overview, the user may have ideas about how to change a model to improve its performance.  They may want to add layers or switch out problematic layers.  
Using the Variations tab in \syss, seen in Figure~\ref{fig:variations}, users can sample new models that are similar to the baseline model.
By default, the variation command will randomly remove, add, replace, or reparameterize layers. 
% However, users may have gained some insights into which types of variations are most promising, either through the ablation studies or their exploration of other models via the model overview.  
Users can constrain the random generation of variations by specifying a subset of types of variations for a given layer.  For example, a user might not want to remove or replace a layer that was very important according to the ablation studies, but could still allow it to be reparameterized.  Valid variation types are \textit{prepend} with a new layer, \textit{remove} a layer, \textit{replace} a layer, or \textit{reparameterize} a layer.

When generating ablations and variations, the user is shown each child model generated from the baseline model that is selected (Fig.~\ref{fig:interface_overview}C).  Changes that were made to generate that model are shown as well.  By viewing all children on the same table, the user may be able to see the effect of certain types of changes; e.g. adding a dense layer typically dramatically increases the number of parameters, while adding a convolutional layer early sometimes increases the validation accuracy.  Spark lines communicate the loss curve of each child model as it trains. Each child model is embedded into the model overview, and can be moved to the model drawer to become a model baseline.
% Then, it can have variations done on it, allowing analysis to iterate from child model to grandchild model.

\subsection{Sequential Neural Architecture Chips}

\begin{figure*}[thb]
\centering
%\subcaptionbox{\label{fig:snac}}{\includegraphics[width=0.34\textwidth]{figs/archip}}%
\subcaptionbox{\label{fig:snac}}{\includegraphics[width=0.34\textwidth]{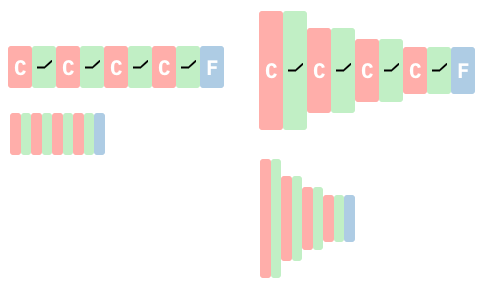}}%%\bigskip
\quad \quad%\hfill
%\par\bigskip
%\vline    \hspace{0.1cm}
\subcaptionbox{\label{fig:projections}}{\includegraphics[width=0.62\textwidth]{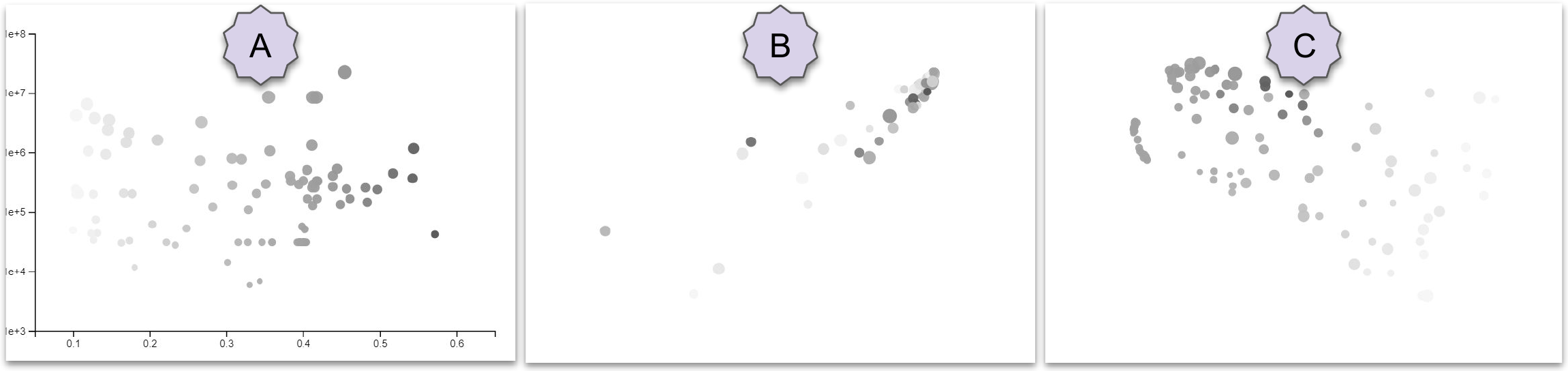}}%
\caption{(a) The \textit{SNAC} visual encoding of a neural network architectures, seen at four different resolutions.  This architecture has a three convolutions, each followed by an activation, and concludes with a fully connected layer. (b) Three alternative visual overviews of the model space.  Section A shows the set of models on a set of interpretable axes, validation accuracy vs. log of the number of parameters.  Sections B and C  use multidimensional scaling to lay out the same set of models based on structural similarity (B) and prediction similarity (C).  The darkness of the circle encodes the model accuracy, and the radius of the circle encodes the log of the number of parameters.}
\end{figure*}

We developed a visual encoding, SNAC (Sequential Neural Architecture Chip), for displaying sequential neural network architectures.  Seen in Figure~\ref{fig:snac}, SNAC is designed to facilitate easy visual comparisons across several architectures via juxtaposition in a tabular format.  Popular visual encodings used in the machine learning\cite{zeiler2014visualizing,krizhevsky2012imagenet,lecun1995convolutional,he2016deep,szegedy2015going} and visual analytics literature\cite{tzeng2005opening,wongsuphasawat2018visualizing,Kahng2016} take up too much space to fit multiple networks on the same page.  In addition, the layout of different computational components and the edges between them makes comparison via juxtaposition difficult\cite{gleicher2018considerations}.

% Early visualizations of network architectures showed the full detail of the network's topology, as neural networks were typically only a handful of layers\cite{tzeng2005opening}.  As time went on and the size of networks increased to have around five to fifteen layers, neurons were no longer shown and each layer was represented by boxes corresponding to the size of the parameters\cite{zeiler2014visualizing,krizhevsky2012imagenet,lecun1995convolutional}.  In recent years, as the number of layers moved into the dozens, networks were represented by computational graphs with each layer as a node and each connection as an edge\cite{he2016deep, szegedy2015going}.  None of these existing encodings allowed for dozens of networks to be visualized on the same page, and the layout of the different networks and edges between them makes comparison via juxtaposition difficult\cite{gleicher2018considerations}.  This drove us to develop our own visual encoding.

The primary visual encoding in a SNAC is the sequence of types of layers.  This is based on the assumption that the order of layers is displayed in most other visualizations of networks.  Layer type is redundantly encoded with both color and symbol.  Beyond the symbol, some layers have extra decoration.  Activation layers have glyphs for three possible activation functions: hyperbolic tangent ($tanh$), rectified linear unit (ReLU), and sigmoid.  Dropout layers feature a dotted border to signify that some activations are being dropped.  \strikeg{The second type of data encoded by SNAC is the size of the data flowing through the network, since this is the second most frequently encoded piece of information in existing visual encodings of network architectures.  }The height of each block corresponds to the data size on a bounded log scale, to indicate to the user whether the layer is increasing or decreasing the dimensionality of the activations flowing through it.  SNACs will be released as an open source component for use in publications and visual analytics tools.\footnote{The open source implementation of SNACs can be viewed at \url{http://www.eecs.tufts.edu/~dcashm01/snacs/}}.

% \begin{figure}
%     \centering
%     \includegraphics[width=0.5\textwidth]{figs/adam/adam-queue.pdf}
%     \caption{The Queue tab shows the user what the backend is currently training, along with estimates of how long it will take to complete training.  Users can reorder the queue, or remove models from it.}
%     \label{fig:queue}
% \end{figure}

% As new models are spawned, they are moved into a training queue
% , represented as a dynamic table, seen in Figure~\ref{fig:queue}.  
% Users can view the training queue and reorder it if they want to prioritize certain models.  They can also remove models from the list of models to be trained.  The queue also displays spark lines representing the training curves of each trained model.  The spark lines can quickly cue the user to models of interest as they scan through the list of models trained during their analysis.

\section{Expert Validation Study}

% \dylan{Is it appropriate to call this a "two stage design study" rather than a design study and an expert validation study?  I don't know if the term "validation study" means we have to report on quantitative outcome.  I also don't know if we can change our terminology at this point.  }

The initial version of \sys was developed based on a design study described in section~\ref{sec:design_study}.  Two months later, a validation study was held with the same four model builders that participated in the design study.  \add{The goal of the validation study was to assess whether the features of \sys were appropriate and sufficient to enable a semi-automated model search, and to determine if the system aligned with the mental model of deep learning model builders.} Users were asked to complete two tasks using \syss, and then provide feedback \add{on how individual features supported them in their tasks}.

\vspace{3pt}
\noindent\textbf{Participants:} The same four individuals with experience in designing deep neural network architectures that participated in the first study agreed to participate in the validation study.  Participants were compensated with a forty dollar gift card.

\vspace{3pt}
\noindent\textbf{Method:}  
Interviews were again held one-one-one using an online conferencing software and took approximately two hours each.  Audio of the conversation as well as screen sharing were recorded. 

At the start of the study, participants were first given a short demo of the system, with the interviewer sharing their screen and demonstrating all of the features of \syss.  Then, participants were given access to the application through their browser and were given two tasks to complete using the tool.  The participant's screen was recorded during their completion of the two tasks.  Participants were asked to evaluate the features of the tool through their usage in completing their tasks.  One out of the four participants was unable to access the application remotely, and as a result, directed the interviewer on what interactions to make in \sys and followed along as the interviewer shared their screen.  

Both tasks consisted of discovering a performant neural network architecture for image classification on the CIFAR-10 dataset, a collection of 50,000 training images and 10,000 testing images each labeled as one of ten mutually exclusive classes\cite{krizhevsky2009learning}.  This dataset was chosen because all four experts \replace{were familiar with it}{had experience building neural network architectures for this dataset.  This allowed the participants to quickly assess whether the system enabled them to do the types of operations they might have done manually searching for an architecture on CIFAR-10.  The CIFAR-10 dataset is perhaps the most widely-studied dataset for deep learning, and competitive baselines found in the literature achieve very high accuracy, well above 90\%\cite{he2016deep}.  In this evaluation, we do not report the performance of the models discovered by participants in this study, but rather their feedback on whether the tool enabled them to navigate the model space in a similar manner to their manual model discovery process.}

% Although it is considered a ``solved'' dataset in that many architectures exist that are able to perform at or above human abilities at classifying its images, it was chosen because it is so familiar that the experts would have considerable domain knowledge of the dataset \remco{repetitive}.  They would be aware of which classes were difficult and how the dataset was composed and balanced.  

\vspace{3pt}
\noindent\textbf{Tasks:}  
The first task given to the participants was to simply find the neural network architecture that would attain the highest accuracy on the 10,000 testing images of CIFAR-10.  For the second task, participants were given a scenario that dictated constraints on the architecture they had to find.  Participants were asked to find a neural network architecture for use in a mobile application used by bird watchers in a certain park that had many birds and many cats.  Birds and Cats are two of the ten possible labels in the CIFAR-10 dataset.  The resulting architecture needed to prioritize high accuracy on those two labels, and also needed to have under 100,000 parameters so that it would be easily deployable on a mobile phone.  The two tasks were chosen to emulate two types of usage for \syss: unconstrained model search and constrained model search.

Participants were given up to an hour to complete the two tasks and were encouraged to ask questions and describe their thought process.  Then, they were asked about the efficacy of each feature in the tool.

% \remco{this part is a little weird in that you don't show what the initial version of the system looks like. As a result, a reader doesn't learn much of anything from your effort of doing this study as iterative design. If space permits, I would expand on this similar to the way that we discussed Snowcat. Along this line, I would break up this section into two sections -- one for each round of interview. This way your writing doesn't ping-poing from talking about the two rounds of interviews, then back to talking about the finding from the first.}

\vspace{3pt}
\noindent\textbf{Findings:}
\add{Participants were able to select models for both tasks.  However, each participant expressed frustration at the lack of fine-grained control over the model building process.}  \strikeg{All four participants offered qualitative feedback on the tool in their second interview.}  In general, participants found that the tool could be useful as an educational tool for non experts because of the visual comparison of architectures.  They also acknowledged that using the tool would save them time writing code to run fine tuning experiments.  \add{We describe participant feedback on individual features of the system and then outline two additional features added to address these concerns.} 

\subsection{Participant Feedback}
\noindent\textbf{Model overview: }
All participants made extensive use of the model overview with interpretable axes, seen in Figure~\ref{fig:projections}(a), to find baseline models.  Two participants started by selecting the model with the highest accuracy irrespective of parameter size, while one participant selected smaller models first, noting that they start with smaller models when they manually select architectures: \textit{"My intuition is to start with simple models, not try a bunch of random models, using your model overview."}  Another participant noted that rather than start with the model with the highest accuracy, they \textit{"thought it would make more sense to find a small model that is doing almost as well and then try to change it."}

Two participants appreciated using the model overview based on prediction similarity.  One noted "\textit{To me, exploring the models in that space seems like a very appealing thing to do.  ... To be able to grab a subselection of them and be able to at a glance see how they are different, how do the architectures differ?}".  
Another participant used the model view in trying to find a small architecture for the second task that performed well on cats and birds: "\textit{instead of looking at every model, I start with a model good at birds, then look at prediction similarity.  Since it does good on birds, I'm assuming similar models do well on birds as well}".  That participant explored in the neighborhood of their baseline model for a model that also performed well on cats.

% When working on the second task, all participants used the data selector to make the projected view display accuracy on the important classes, \textit{bird} and \textit{cat}.  

\noindent\textbf{Model drawer and inspection: }
Each participant moved multiple interesting models into the model drawer, and then inspected each model in the model inspector.  They all used the confusion matrix to detect any poor qualities about models.  Several participants ignored or discarded models that had all zeros in a single row which indicated that the model never predicted an instance to be that class across the entire testing dataset.

% , judging that there was something endemic with those models.

\noindent\textbf{Generation of new models: }
While some participants found the ablation studies interesting, one participant noted that some ablations were a waste of resources: "\textit{I basically don't want my system to waste time training models that I know will be worse... For example, removing the convolutional layer.}".  Some participants used their own background and experience to inform which variations they did, while others used the model overview and model drawer to discover interesting directions to do variations in.  When viewing two architectures with similar accuracy but very different sizes, a participant commented "\textit{I can visually tell, the only difference I see is a pink color.  It's a nice way to learn that dense layers add a lot of parameters.}"

% \dylan{Quote about switching out pool layers, appending layers}.  For the second task, one participant found one model that performed well on the  \textit{bird} class, and one model that performed well on the \textit{cat} class, and tried to generate variations between them.  \dylan{Quote about bird vs. cat}.  

All participants expressed a desire to have more control over the construction of new models.  This would allow them to do more acute experimentation once they had explored in the neighborhood of an interesting baseline model. One participant described it as the need for more control over the model generation process: "\textit{I think we need more customization on the architecture.  Currently, everything is rough control ... Of course for exploring the search space, rough control would be more helpful.  But for us to understand the relation [between architecture and performance], sometimes we need precise control.}"  \add{All participants noted that relying on rough control resulted in many models being spawned that were not of interest to them, especially once they had spent some time exploring the model space and knew what kind of model they wanted to generate.}

% Participants also suggested that there needed to be better 

% Two participants were surprised at the type of models that were initially generated, and suggested a less random search.  They suggested that the randomness of the search made it difficult to develop any understanding of the space.  They recommended something more like a beam search, where the location of the convolutional layers and fully connected layers were more controlled, because those were the most important layers.

\subsection{System Updates}

The feedback from the expert validation study led to two changes to the system.  \add{Both changes allow for more fine-grained control over which models were generated, both to allow for more precise experimentation and to reduce the number of models that need to be trained.}

\begin{figure}
    \centering
    \includegraphics[width=0.5\textwidth]{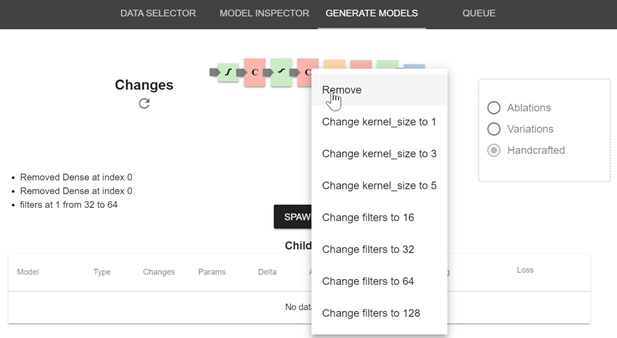}
    \caption{The ability to handcraft models was added based on feedback from a validation study with model builders.  Starting from a model baseline, users can remove, add, or modify any layer in the model by clicking on a layer or connections between layers.  This provides fine-grained control over the models that are generated.}
    \label{fig:handcrafted}
\end{figure}

% \begin{figure}
%     \centering
%     \includegraphics[width=0.5\textwidth]{figs/adam/adam-ablations.pdf}
%     \caption{Running all ablations took too much time according to the validation study, so we added a selection brush, controlling which ablations are run.}
%     \label{fig:brushed_ablations}
% \end{figure}

\begin{itemize}[itemsep=0pt, topsep=3pt, partopsep=3pt, leftmargin=9pt]
    \item \textbf{Change C1: Creating Handcrafted Models:} While variations proved useful for seeing more models in a small neighborhood in the model space, participants expressed frustration at not being able to explicitly create particular architectures.  To address this, we added the handcrafted model control, seen in Figure~\ref{fig:handcrafted}.  Users see the same SNAC used in the Ablations and Variations controls, but with additional handles preceding each layer.  By clicking on the layer itself, users can select to either remove a layer or reparameterize it.  By clicking on the handles preceding each layer, the user can choose to add a layer of any type.  
\item \textbf{Change C2: Subselections of ablations:}  Two participants found that the ablations tool wasted time by generating models that weren't particularly of interest to the user.  We added a brushing selector, seen in Figure~\ref{fig:ablation} to allow the user to select \textit{which} layers were to be used in ablations, so that the user could quickly run ablations on only a subset of the model.
    % \item \textbf{Change C3: Better attribution of child models:}  During experimentation, found it difficult to remember the relationships between models that were generated.  While there were links between parents and children, since children are shown in a table beneath a parent as in Figure~\ref{fig:interface_overview}(c), when viewing a child in the model inspector (Figure~\ref{fig:model_inspection}) or the queue tab~\ref{fig:queue}, there was no visible link to the parent.  As a result, we added links to the parents in both places, as well as information into the queue about what type of change (ablation, variation, handcrafted) resulted in the birth of that model.
\end{itemize}

% In our interview studies, two participants gave feedback on an initial version of variations that they wanted to explicitly give exact tweaks to a mdoel, and thus suggested that we give the user an exact copy of the model to then modify.  This resulted in the "copy" feature, which would allow users to reorder layers, add/remove layers, and explicitly reparameterize layers to produce exact architectures.  

% Second, all participants noted that the training time for models takes a very long time, making it difficult to do in situ analysis, although they noted that this is a problem in manual neural architecture search as well.  Two participants suggested that the system could provide more feedback on which parts of an architecture lead to more parameters and thus more training time.  One participant also pointed out that it would be helpful to be able to kill the current training process, if the system starts training a very large model.

\section{Use Case: Classifying Sketches}

To validate the new features suggested by the study, we present a use case for generating a performant, small model for an image classification dataset.  In this use case, we refer to tasks T1, T2, and T3 supported by our system as outlined in section~\ref{sec:design_study}.

Leon is a data scientist working for a non-governmental organization that researches civil unrest around the world.  He is tasked with building a mobile app for collecting and categorizing graffiti, and would like to use a neural network for classifying sketched shapes.  Because his organization would like to gather data from all over the world, the application must be performant on a wide swath of mobile devices.  As a result, he needs to consider the tradeoff between model size and model accuracy.

\vspace{3pt}
\noindent\textbf{Data:}  
He downloads a portion of the \textit{Quick Draw} dataset to use as training data for his image classifier.  Quick Draw is a collection of millions of sketches of 50 different object classes gathered by Google\cite{quickdraw}.  Rather than download the entire dataset, Leon downloads 16,000 training images and 4,000 training images from each of 10 classes that are commonly found in graffiti to serve as training data\footnote{For this use case, we used the 10 most convergent classes in Quick Draw as identified by Strobelt et al.\cite{ffdata}}.  Overnight, he uses \sys to auto-generate an initial set of 100 models, and the next day, he loads up \sys to begin his model search.

\begin{figure}
    \centering
    \includegraphics[width=0.5\textwidth]{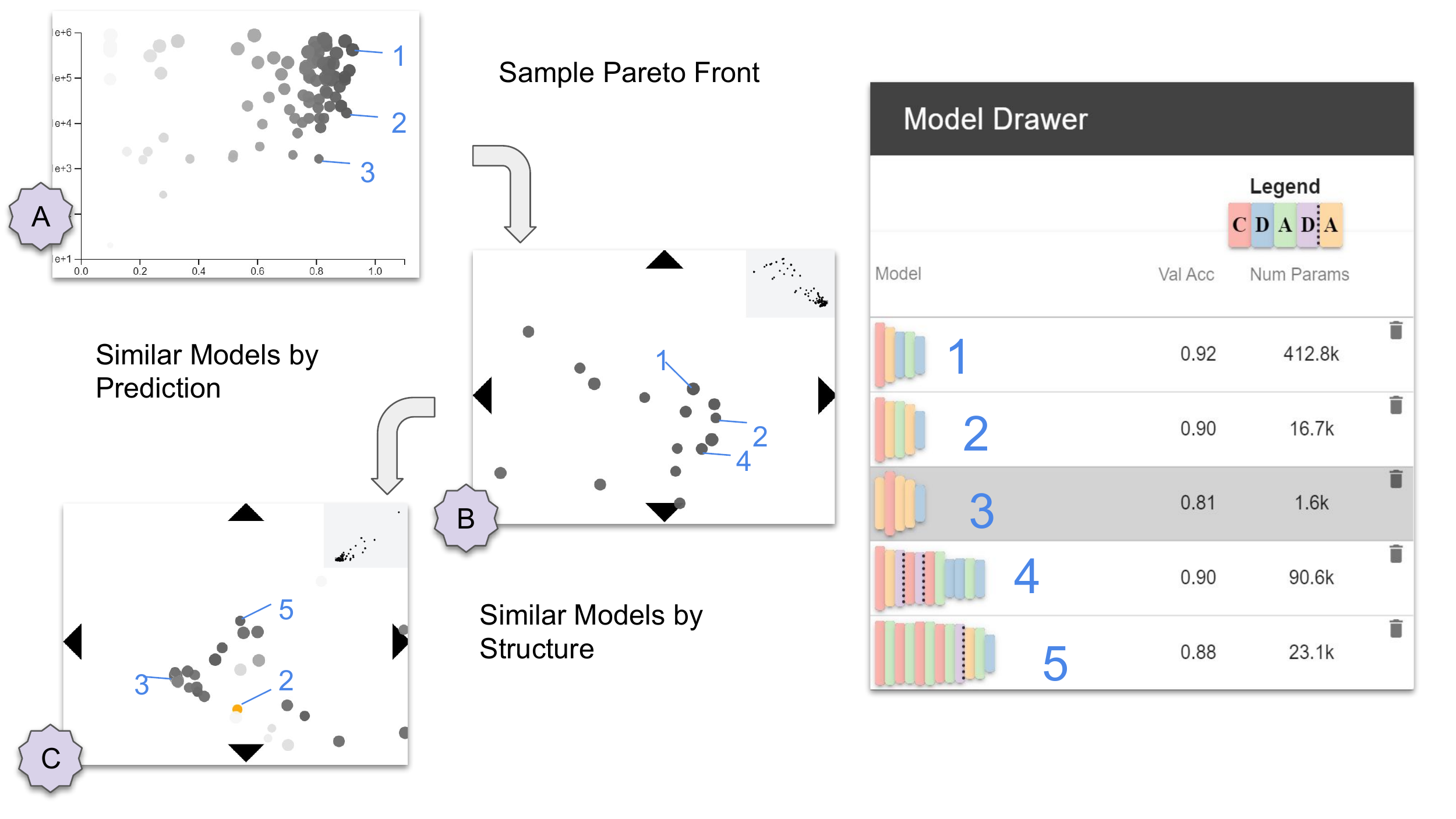}
    \caption{In our use case, the model builder first samples models 1, 2, and 3 on the pareto front of accuracy vs. model size.  He then selects models 4 and 5 from the two alternative model overviews provided.}  
    \label{fig:use_case_model_overview}
\end{figure}

\vspace{3pt}
\noindent\textbf{Search for baselines in the model overview:}  
To find a set of baseline models [T1], he starts with the default model overview, seen in Figure~\ref{fig:use_case_model_overview}A.  He sees that there are many models that achieve at or above 90\% accuracy, but they appear to have many parameters.  He samples three models from the pareto front, two which have the high accuracy he desires and one which has an order of magnitude less parameters.  He switches the model view to lay out models based on performance prediction similarity (Figure~\ref{fig:use_case_model_overview}B) and hovers the mouse around the neighborhood of his selected models to see what alternative architectures could result in similarly good performance, and adds an additional model which has multiple convolutional and dense layers, as well as some dropout layers.  Lastly, he switches the model view to lay out models based on structural similarity (Figure~\ref{fig:use_case_model_overview}C) to see how small differences in architecture correspond to changes in either accuracy or parameters [T3].  He selects a fifth model which differs from his previously selected models in that it spreads its convolutional filters over multiple layers instead of concentrating them in a single initial layer.

\vspace{3pt}
\noindent\textbf{Ablations:}  
He decides to start with the smallest model, model 3, since it has reasonably high accuracy of 81\% and a very small amount of parameters, approximately 1600.  Having chosen a baseline, he moves on to generate local, constrained searches in the neighborhood of the baseline [T2].  After checking in the model inspector that the model performs reasonably well on all classes, he runs ablations on this model and sees that removing the first and last max pool layers increased both accuracy and the number of parameters.  He notes that, with an accuracy of 90\% and 11.9k parameters, the model resulting from removing the first max pool layer is now on the pareto front between validation accuracy and number of parameters, so he adds it to his model drawer for further consideration.

\vspace{3pt}
\noindent\textbf{Variations:}  
While the ablations indicated that he may want to remove some of the pooling layers, he wants to see the effects of various other modifications to his baseline model.  He decides to generate variations of all kinds (\textit{prepend, remove, replace, reparameterize}) along the pooling layers, and also allow for reparameterization of the convolutional layer.  He generates 10 new variations from those instructions, and by looking at their results, sees that increasing the number of convolutional filters results in too many parameters, but this can be compensated for by also increasing the pool size.

\begin{figure}
    \centering
    \includegraphics[width=0.45\textwidth]{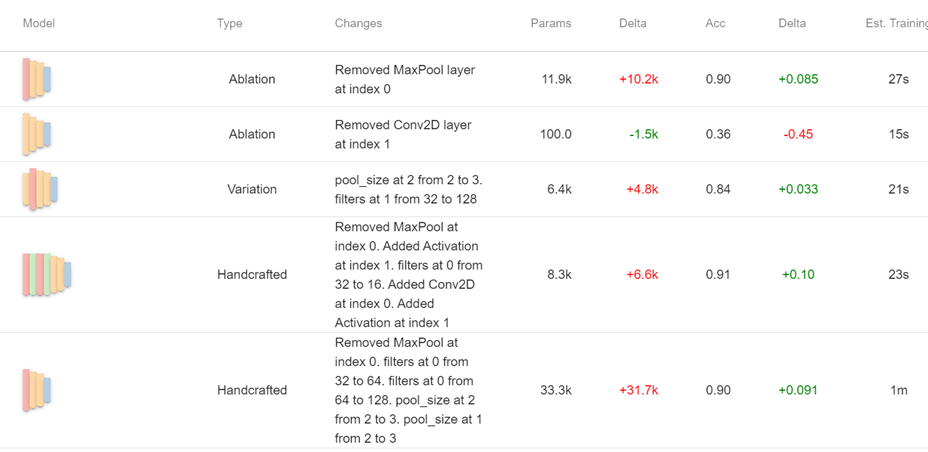}
    \caption{After generating ablations, variations, and several handcrafted models, the model builder compares all discovered models and chooses the model in the fourth row, because of its high accuracy and low number of parameters.}  
    \label{fig:use_case_results}
\end{figure}

\vspace{3pt}
\noindent\textbf{Handcrafting Models:}  
After developing an understanding of the model space, he generates some handcrafted models.  He removes the first max pooling layer because that helped in the ablation studies.  He then creates three new models from this template.  First, he splits the starting convolutional layer into three convolutional layers with fewer filters, to be more like model 5.  He then tries adding dropout, to be more like model 4.  Lastly, he creates a model with activations like model 2, and different options chosen for pooling layers and kernels inspired by the variations.  The trained results can be seen in Figure~\ref{fig:use_case_results}.

\vspace{3pt}
\noindent\textbf{Result:}
Leon eventually decides on using an architecture with 91\% accuracy and only 8.3k parameters, seen in the fourth row of Figure~\ref{fig:use_case_model_overview}.  This model has comparable accuracy to models 1 and 2 that were initially chosen from the pareto front, seen in Figure~\ref{fig:use_case_model_overview}, but drastically fewer parameters than model 1 (412.8k) and model 2 (16.7k).  As a result, the architecture found by Leon can be deployed on older technology and classify images faster than any of the initially sampled models.
% In order to get a better idea of the effect of different layers on performance, 

\section{Discussion}

\subsection{Human in the Loop Neural Architecture Search}

\add{Our experience and study suggested that the presence of a human in the loop benefited the discovery of neural architectures.  However, a common pattern in deep learning research is for applications to start with the neural network as an independent component in a set of semantic modules, only for subsequent research to point out that subsuming all components into the neural network and training it end-to-end results in superior performance.  As an example, the \textit{R-CNN} method for object recognition dramatically outperformed baselines for object detection using a CNN in concert with a softmax classifier and multiple bounding box regressors\cite{girshick2014rich}; however, its performance was eclipsed only one year later by \textit{Fast R-CNN}, which absorbed the classifiers and regressors into the neural network\cite{girshick2015fast, zhao2019object}.  This suggests that the user processes in \sys, such as selecting models on the pareto front and running certain ablations and variations, could be automated, and the whole process run end to end as a single optimization without a human in the loop.  Ultimately, this perspective ignores the tradeoffs that users are able to make; users can very quickly and efficiently narrow the search space to only a small subset of interesting baselines based on a number of criteria that are not available to the automated methods.  These include fuzzy constraints on the number of parameters, a fuzzy cost function that differs per class and instance, and domain knowledge of the deployment scenario of the model.  For this reason, we advocate that the human has a valuable role when searching for a neural architecture using \sys.}

\subsection{Generalizability}

\add{The workflow of \sys is generalizable to other types of automated machine learning and model searches beyond neural networks.  The two primary components of \sys are a set of projections of models and a local sampling method to generate models in a neighborhood of a baseline model.  As long as these two components can be defined for a model space, the workflow of \sys is applicable.  Of the three projections used, both the semantically meaningful projection of accuracy vs. number of parameters and the prediction similarity distance metrics are generalizable to any machine learning model, while structural similarity distances can be easily chosen, such as the euclidean distance between weights for a support vector machine.   Similarly, random sampling in the neighborhood of a model can be done in any number of ways; if the model space is differentiable, gradient-based techniques can be used to sample in the direction of accurate or small models.}

\subsection{Scalability}
In order to facilitate human in the loop neural architecture search, \sys must make several constraints on its model space.  It limits the size of the architectures it discovers so that they can be trained in a reasonable amount of time while the user is engaged with the application.  \strikeg{This is based on two assumptions.  First, there are some users who are willing to use smaller architectures because their performance is still fairly high, and above what non-neural models would be able to get.  Second, for some use cases, smaller models are actually preferred because they can be deployed on smaller devices and are more energy efficient. 

For some users and use cases, however, these assumptions could be invalid. Consider the medical domain, where image classifiers are trained to make diagnoses of various scans and images.  Data scientists might want to use a tool like \sys to discover architectures specific to each image classification task they have; the architecture that is relevant for detecting cancer might be different than the architecture for detecting bone chips.}\add{In certain domains, however, the tradeoff between accuracy and size of the model is very different; stakeholders don't want to sacrifice any accuracy.}  In that case, the cap on model size in \sys could be removed\add{, and \sys could be used to find large networks that take many hours to train}.  \strikeg{We note that our client-server architecture could have its server run on a cloud machine with multiple GPUs to speed up training, but at a certain scale of model size, it is inevitable that training will take a long time.  A user might have to wait for an hour or more before the results of their ablation studies were returned, which only took about three minutes in our use case.}  It isn't feasible to expect a user to stay in situ the entire time while \sys trained the several dozen models needed to enable architecture discovery.  Instead, a dashboard-like experience, easily viewable in a casual setting on a small screen such as a phone might be preferable.  In general, the types of user experiences used in visual analytics tools for machine learning models may have to be adapted to the scale of time necessary for constructing and searching through industry-level neural networks.  These adaptations offer an interesting avenue for future research in user experience.

The visual encoding used for neural network architectures, SNACs, can only display network architectures that are linked lists, which leaves out some newer types of architectures that have \textit{skip connections}, which are additional linkages between layers.  This problem could be solved by improving the encoding to communicate skip connections.  Ultimately, supporting every possible network architecture amounts to supporting arbitrary graphs, and there is no space efficient way to do so without losing information.  For that reason, we limit the scope in this project to network architectures that are linked lists, because they are simpler to understand and are a common architecture that are more performant than non-neural network models for image classification problems. 

\strikeg{There are also visual scaling issues in the number of classes in the data.  While we demonstrate our system with image datasets with 10 classes, it is entirely plausible to have image datasets with thousands of classes\cite{krizhevsky2012imagenet, DBLP:journals/corr/abs-1710-06501}.  Our system displays examples from each class in the data selector to allow the user to view performance on specific classes.  This could be adapted to a larger number of classes by providing a searching mechanism.  Users can also inspect model performance through the use of a confusion matrix, seen in figure~\ref{fig:model_inspection}.  The adaption of a confusion matrix to thousands of classes is still an open avenue of research\cite{dilawer2017classifiers}, and we offer no suggestion other than providing the class-by-class accuracy in a searchable interface.}

\section{Conclusion}

Neural networks can be difficult to use because choosing an architecture requires tedious and time consuming manual experimentation.  Alternatively, automated algorithms for neural architecture search can be used, but they require large computational resources and cannot accommodate soft constraints such as trading off accuracy for model size or trading off on performance between classes.  We present \syss, a visual analytics tool that allows a model builder to discover a deep learning model quickly via exploration and rapid experimentation of neural network architectures and their parameters.  \sys enables users to quickly search for baseline models through a visual overview, visually compare subsets of those models to understand small, local differences in architectures, and then generate local, constrained searches to fine tune architectures.  Through a design study with four model builders, we derive a set of design goals.  We provide a use case in building a small image classifier for identifying sketches in graffiti that is small enough to used on even very old mobile devices.  We demonstrate that the semi-automated approach of \sys allows users to discover architectures quicker and easier than through manual experimentation or fully automated search.

\acknowledgments{
We would like to thank Subhajit Das, who provided code for confusion matrices, and Kirthevasan Kandasamy, who gave assistance incorporating the OTMANN distance metric.  Support for the research is partially provided by DARPA FA8750-17-2-0107 and NSF CAREER IIS-1452977. The views and conclusions contained in this document are those of the authors and should not be interpreted as representing the official policies, either expressed or implied, of the U.S. Government.  
}

\bibliographystyle{abbrv-doi}

\bibliography{template}
\end{document}